\documentstyle[preprint,aps,epsf]{revtex}	
\begin{document}
\thispagestyle{empty}
%
%
\preprint{$
\begin{array}{l}
\mbox{FERMILAB-Conf-98/132-T}\\[-3mm]
\mbox{IASSNS-HEP-98-41}\\[-3mm]
\mbox{BA-98-23}\\[-3mm]
\mbox{April 1998} \\   [2.mm]
\end{array}
$}
\title{Implications of a Minimal $SO(10)$ Higgs Structure\footnote
	{Contribution submitted to the NEUTRINO 98 Conference to 
	be held in Takayama, Japan,\\[-0.1in] \hspace*{0.25in} June 4-9, 1998.}
	\footnote{Email: $^1$albright@fnal.gov,
	$^2$babu@ias.edu, $^3$smbarr@bartol.udel.edu}}
\author{Carl H. Albright$^1$}
\address{\baselineskip 15.pt
        Department of Physics, Northern Illinois University, DeKalb, 
        IL 60115\\
        and \\
        Fermi National Accelerator Laboratory, P.O. Box 500, Batavia, IL 
        60510\\[-0.1in]}
\author{K.S. Babu$^2$}
\address{\baselineskip 15.pt
	School of Natural Sciences, 
        Institute for Advanced Study, Princeton, NJ 08540\\[-0.1in]}
\author{S.M. Barr$^3$}
\address{\baselineskip 15.pt
        Bartol Research Institute, University of Delaware,
        Newark, DE 19716}
\maketitle
\begin{abstract}
\baselineskip 15.pt

A minimal $SO(10)$ Higgs structure involving a single adjoint field along
with spinors, vectors and singlets  has been shown to break the $SO(10)$ 
gauge symmetry to the standard model while stabilizing the F-flat directions 
and solving the doublet-triplet splitting problem naturally.  With this 
minimal set of Higgs fields, we show how to construct quark and lepton mass 
matrices which explain well the many features of the observed spectrum,
including the Georgi-Jarlskog mass relations.  A large $\nu_{\mu} - 
\nu_{\tau}$ mixing angle results naturally as observed in the atmospheric 
neutrino data.  A particular model relying on a family symmetry
has been constructed which realizes the desired mass matrices.
\end{abstract}
\newpage
\indent A brief discussion is given of the implications of a minimal $SO(10)$ 
Higgs structure that have been developed in a recent series of papers.  Barr 
and Raby \cite{BR} have shown how this minimal set of Higgs fields breaks the 
$SO(10)$ gauge symmetry to the standard model while stabilizing the F-flat 
directions and thus solves the double-triplet splitting problem.  Following 
this lead, the authors \cite{ABB} have used this Higgs
structure to construct quark and lepton mass matrices which are fairly
tightly constrained with some interesting features emerging.  Of special
interest to this Conference is the large $\nu_{\mu} - \nu_{\tau}$ mixing
angle resulting from the special textures of the Dirac matrices, as opposed 
to the more conventional large hierarchical structure for the 
Majorana neutrino matrix \cite{Maj}.

\section{Minimal Higgs Structure}

We begin with a summary of the minimal $SO(10)$ Higgs structure \cite{BR} 
which solves the doublet-triplet splitting problem naturally rather than
by fine-tuning.  The Higgs fields 
which are involved consist of a pair of ${\bf 10}$'s, one ${\bf 45}$,
two pairs of ${\bf 16} + {\bf \overline{16}}$'s and four singlets.  
The Higgs superpotential is written 
$$\begin{array}{rcl}
	W &=& T_1 A T_2 + M_T T_2^2 + W_A + W_C + W_{CA} + W_{TC}
		\nonumber \\[0.1in]
	W_A &=& {\rm tr} A^4/M + M_A {\rm tr} A^2\\
	W_C &=& X(\overline{C} C)^2/M_C^2 + f(X)\\
	W_{CA} &=& \overline{C}'(P A / M_1 + Z_1) C +
		\overline{C} (P A/M_2 + Z_2) C'\\
        W_{TC} &=& \lambda T_1\overline{CC}\\
\end{array} \eqno(1)$$
Here $T_1$ and $T_2$ label the two ${\bf 10}$'s, $A$ labels the ${\bf 45}$,
$C,\ \overline{C},\ C',\ \overline{C}'$ label the two pairs of ${\bf
16} + {\bf \overline{16}}$'s, while $P,\ X,\ Z_1,\ Z_2$ label the four
singlets.

The $W_A$ terms produce the Dimopoulos - Wilczek mechanism \cite{DW} by 
generating
a VEV for the single ${\bf 45}$ in the $B - L$ direction.  The $T_1 A T_2$
term gives superheavy masses to the color triplets in $T_1$ and $T_2$.
The mass term $M_T T^2_2$ gives superheavy masses to the $T_2$ doublets
as well.  As a result of the presence of $W_C$, the $F_X = 0$ condition
forces the $C$ and $\overline{C}$ pair to get VEVs in the $SU(5)$-singlet
direction.  The VEVs of $A$ and $C$ then break $SO(10)$ to the standard
model.  The term $W_{CA}$ couples $C$ and $\overline{C}$ to $A$ and 
prevents the production of colored pseudo-goldstone bosons in the 
breaking of $SO(10)$.  Since no GUT-scale VEVs are generated for $C'$
and $\overline{C}'$, the Dimopoulos - Wilczek hierarchical form of 
$\langle A \rangle$ is not destabilized by the presence of $W_{CA}$, thus
solving the doublet-triplet splitting problem.  Finally, the presence of 
the term $W_{TC}$ induces an electroweak breaking VEV for $C'$ which 
mixes with that in $T_1$.  Hence the two Higgs doublets appear in the 
combinations
$$\begin{array}{rcl}
H &=& {\bf 5}(T_1)\nonumber\\
H' &=& {\bf \overline{5}}(C')\cos \theta  - {\bf \overline{5}}(T_1) \sin 
	\theta.
\end{array} \eqno(2)$$
in terms of the $SU(5)$ representations present in $T_1$ and $C'$.  The 
combination orthogonal to $H'$ gets massive and drops out of the picture.

An important point to be made is that the above form of the Higgs 
superpotential can be uniquely obtained by the introduction of a $U(1) 
\times Z_2 \times Z_2$ family symmetry \cite{BR} with the 
appropriate assignment for the charges of the Higgs fields as follows:

$$\begin{array}{rllll}
        &A(0^{+-}), & T_1(1^{++}), & T_2(-1^{+-})  \\[0.1in]
        &C({1\over{2}}^{-+}),\quad & \overline{C}(-{1\over{2}}^{++}),\quad & 
                C'(\left[{1\over{2}}-p\right]^{++}),\quad & \overline{C}'
                (\left[-{1\over{2}}-p\right]^{-+}) \\[0.1in]
        &X(0^{++}), & P(p^{+-}), & Z_1(p^{++}), & Z_2(p^{++})
\end{array} \eqno(3)$$

\section{Fermion Mass Matrices from the Minimal Set of Higgs Fields}

We can then attempt to construct fermion mass matrices from the VEVs 
appearing in the minimal set of Higgs fields.  The VEVs in question appear
at the GUT scale and at the electroweak scale as follows:
	$$\begin{array}{rl}
		\Lambda_G:& \qquad \langle A \rangle,\ \langle C \rangle,\ 
		\langle \overline{C} \rangle,\ \langle P \rangle,\ 
		\langle X \rangle,\ \langle Z_1 \rangle,\ \langle Z_2 
			\rangle  \\
		\Lambda_{ew}:& \qquad \langle T_1 \rangle,\ 
		\langle C' \rangle \\
	  \end{array} \eqno(4)$$

\noindent Note that since the VEVs of the doublets of the $T_1$ $SO(10)\ 
{\bf 10}$ appear in the $SU(5)\ {\bf 5 + \overline{5}}$ pair, $\langle
T_1 \rangle$ couples symmetrically in family space to all members of 
a pair of ${\bf 16}$ fermions, whether up or down quarks, neutrinos or charged
leptons.  On the other hand, since the $C'$ VEV of the doublet appears only
in the $SU(5)\ {\bf \overline{5}}$ of the ${\bf 16}$, this VEV couples only
to the down quarks and charged leptons in a ${\bf 16}$ and ${\bf 10}$ 
fermion pair and asymmetrically at that.  All the GUT scale VEVs except
$\langle A \rangle$ are $SU(5)$ 
singlets, with $\langle A \rangle$ of the single $SO(10)\ {\bf 45}$ assigning 
an antisymmetric $B - L$ quantum number of magnitude $1/3$ or $1$ to the 
quarks and leptons, respectively. 

Yukawa coupling unification at the GUT scale suggests as usual the 
coupling of $\langle T_1 \rangle$ to the third generation quarks and leptons
according to ${\bf 16}_3 {\bf 16}_3 T_1$.
Now, however, because of the linear combination appearing in (2), the
top-to-bottom quark mass ratio at the GUT scale assumes the form:
	$$m^o_t/m^o_b = \tan \beta / \sin \theta \eqno(5)$$
in terms of the $\langle {\bf \overline{5}}(T_1) \rangle - \langle 
{\bf \overline{5}}(C') \rangle$ mixing angle $\theta$.  Hence $\tan \beta$
can assume any value in the range 2 - 55.

The Georgi-Jarlskog relations \cite{GJ}, $m_s^0 \cong m_{\mu}^0/3$ and $m_d^0 
\cong 3m_e^0$, together with the minimal Higgs structure then suggest the 
following textures for the Dirac mass matrices \cite{ABB}:
$$\begin{array}{ll}
U^0 = \left( \begin{array}{ccc} 0 & \sigma & 0 \\ \sigma & 0 & \epsilon/3 \\
0 & - \epsilon/3 & 1 \end{array} \right) m, & \;\;\;
N^0 = \left( \begin{array}{ccc} 0 & \sigma & 0 \\ \sigma & 0 & - \epsilon \\
0 & \epsilon & 1 \end{array} \right) m, \\ \\
D^0 = \left( \begin{array}{ccc} 0 & \sigma + \sigma' & 0 \\
\sigma + \sigma' & 0 & \rho + \epsilon/3 \\
0 & -\epsilon/3 & 1 \end{array} \right) \tilde{m}, & \;\;\;
L^0 = \left( \begin{array}{ccc} 0 & \sigma + \sigma' & 0 \\
\sigma + \sigma' & 0 & - \epsilon \\ 0 & \rho + \epsilon  & 1
\end{array} \right) \tilde{m},
\end{array} \eqno(6)$$

\noindent where the matrices are written so that the left-handed antifermions
multiply them from the left and the left-handed fermions from the right.
The 2 - 3 sector of the above matrices is essentially uniquely determined.
Here the $\epsilon$ terms arise from the $B - L$ VEVs, $\langle A \rangle$, 
of the antisymmetric ${\bf 45}$, while the $\rho$ terms arise from the $\langle
C' \rangle$ VEV.
The 1 - 2 sector has more uncertainty.  We have made the simplest
choices here; for example, the $\sigma$ terms may arise from $\langle
T_1 \rangle$ Higgs VEVs after integrating out superheavy ${\bf 16}$
fermions, while the $\sigma'$ terms appear after integrating out superheavy
${\bf 10}$ fermions.  

If we assume that $\rho \gg \epsilon \gg \sigma' \gg \sigma$, by diagonalizing
the matrices we find:
$$\begin{array}{l}
m_b^0/m_{\tau}^0 \cong 1 - \frac{2}{3} \frac{\rho}{\rho^2 + 1}
(\epsilon \cos \alpha), \\ \\
m_{\mu}^0/m_{\tau}^0 \cong \epsilon \frac{\rho}{\rho^2 + 1}
\left( 1 - \frac{\rho^2 - 1}{\rho (\rho^2 + 1)}(\epsilon \cos \alpha)
\right), \\ \\
m_s^0/m_b^0 \cong \frac{1}{3} \epsilon \frac{\rho}{\rho^2 + 1}
\left( 1 - \frac{1}{3}
\frac{\rho^2 - 1}{\rho (\rho^2 + 1)}(\epsilon \cos \alpha)
\right), \\ \\
m_c^0/m_t^0 \cong \epsilon^2 /9, \\ \\
V_{cb}^0 \cong \frac{1}{3} \epsilon \frac{\rho^2}{\rho^2 + 1}
\left( 1 + \frac{2}{3} \frac{1}{\rho (\rho^2 + 1)} (\epsilon \cos \alpha)
\right), \\ \\
m_d^0/m_e^0 = 3(1+{2 \over 3 \rho} \epsilon {\rm cos}\alpha), \\ \\
|V_{us}^0| = |\sqrt{m_d^0 \over m_s^0}
	{1 \over (\rho^2+1)^{1/4} } - \sqrt{m_u^0 \over m_c^0}e^{i\phi}|, \\ \\
|V_{ub}^0| \simeq |\sqrt{m_d^0 \over m_s^0} {m_s^0 \over m_b^0}
	{\rho \over (\rho^2+1)^{1/4} } - \sqrt{m_u^0 \over m_c^0} e^{i\phi}
	(\sqrt{m_c^0 \over m_t^0} - {m_s^0 \over m_b^0}{1\over\rho})|. \\ \\
\end{array} \eqno(7)$$

\noindent
Here $\alpha$ is the relative phase between $\epsilon$ and $\rho$, while 
$\phi$ is the relative phase between $\sigma$ and $\sigma'$.  In addition
to the Georgi-Jarlskog relations \cite{GJ},
we observe that $m^0_b \simeq m^0_{\tau}$; $V^0_{cb},\ m^0_{\mu}/m^0_{\tau}$
and $m^0_s/m^0_b \sim O(\epsilon)$; while $m^0_c/m^0_t \sim O(\epsilon^2)$.

Of special interest is the issue of neutrino masses and mixings. 
The light neutrino mass matrix is given by $M_{\nu} = - N^TM^{-1}_RN$,
in terms of the Dirac neutrino matrix and the superheavy right-handed 
Majorana neutrino mass matrix.  If we simply take $M_R$
diagonal and similar to the identity matrix, a large mixing emerges by 
virtue of the form of the Dirac matrices $N^0$ and $L^0$ in Eq. (6) as
indicated below.  In fact, the mixing will generally be very large, unless
the form of $M_R$ is fine-tuned.  As a result of the asymmetrical
$\rho$ contributions appearing in $D^0$ and $L^0$, we can then understand 
why $V_{cb}$ mixing is small in the quark sector while the $\nu_{\mu} - 
\nu_{\tau}$ mixing is large in the neutrino sector.  The atmospheric 
anomaly \cite{atm} can thus be understood without resorting to a very 
hierarchical form for the Majorana matrix.  

\section{Numerical Results}

In order to obtain numerical comparisons with experiment, the fermion 
masses and mixings have been evolved \cite{ABB} from the unification scale, 
$M_G$, to the 
supersymmetry scale $M_{SUSY} \sim m_t$, by making use of 2-loop MSSM $\beta$ 
functions and from $M_{SUSY}$ to the running mass scales with the use 
of 3-loop QCD and 1-loop QED or EW beta functions.  We find the known 
quark mass and mixing data is best fitted with $\tan \beta \simeq 30$.
For this value, and the known $m_{\mu},\ m_{\tau}$ and $V_{cb}$, the 
two parameters $\rho$ and $\epsilon$ are found to be 
	$$\rho = 1.73(1 - \Delta_{cb}),\qquad \epsilon = 0.136(1 - 
		0.5\Delta_{cb}), \eqno(8)$$
in terms of the chargino loop correction $\Delta_{cb} \simeq -0.05$ for 
$V_{cb}$.  

The following predictions then emerge with $\cos \alpha = 1$:
\begin{itemize}
\item	Good agreement with the experimental value
	for $m_b(m_b) = 5.0(1 + \Delta_b)$ GeV is reached with the 
	combined gluino and chargino loop correction $\Delta_b \cong -0.15$.

\item	With $\Delta_s \simeq \Delta_b \cong - 0.15$, $m_s(1 GeV) = 176(
	1 + \Delta_s) = 150$ MeV compared with $180 \pm 50$ MeV.

\item	We find $m_c(m_c) = (1.05 \pm 0.11)(1 - \Delta_{cb}) \sim (1.10 \pm 
	0.11)$ GeV, in reasonable agreement with the experimental value of 
	$(1.27 \pm 0.1)$ GeV.

\item	For a non-hierarchical diagonal form for $M_R$, we find 
	$\sin^2 2\theta_{\mu\tau} \simeq 0.7$.  This large neutrino
	mixing occurs not because of a hierarchy in the right-handed 
	Majorana neutrino mass matrix but rather because of the asymmetrical
	form appearing in the charged lepton mass matrix as a result of the
	minimal Higgs structure assumed.

\item	For the form of the first generation contributions to the mass 
	matrices given in (6), acceptable results for $|V_{us}|$ and 
	$|V_{ub}|$ emerge with the phase $\phi \sim 180^o$.  The
	leptonic mixings $|(U_{\nu})_{e\nu_2}|$ and $|(U_{\nu})_{e\nu_3}|$
	are small and consistent with the small angle MSW solution for 
	the solar neutrinos, but their precise values are sensitive to the 
	assumed structure of $M_R$.  
\end{itemize}

\noindent In \cite{ABB}, detailed results have been obtained for a broader range
of the input parameters $\rho,\ \epsilon,\ \cos \alpha$ and $\phi$.

\section{Specific $SO(10)$ Supersymmetric Grand Unified Model}

It is of interest to construct a specific $SO(10)$ supersymmetric grand 
unified model
which leads to the textures for the mass matrices postulated in Eq. (6).
This has been accomplished in \cite{ABB} for the second and third generation
contributions which are essentially uniquely determined.  The first
generation contributions, being higher order, are less well determined
and are subject to further study as are the contributions to the 
right-handed Majorana matrix.

Considering only the second and third generations, we are led to the 
following Yukawa superpotential,

$$\begin{array}{rcl}
W_{Yukawa} & = & {\bf 16}_3 {\bf 16}_3 T_1 \nonumber\\[0.1in]
& + & {\bf 16} {\bf \overline{16}} P + {\bf 16}_3 {\bf \overline{16}} A
+ {\bf 16}_2 {\bf 16} T_1 \nonumber\\[0.1in]
& + & {\bf 10} {\bf 10}' \overline{C} C/M_P + {\bf 16}_2 {\bf 10} C
+ {\bf 16}_3 {\bf 10}' C'.
\end{array} \eqno(9)$$

\noindent In addition to the two light fermion families, one pair of ${\bf 16}
+ {\bf \overline{16}}$ and one pair of ${\bf 10} + {\bf 10'}$ fermions have 
been introduced which get superheavy as a result of the interactions 
present in Eq. (9).  By making use of the previous $U(1) \times Z_2 
\times Z_2$ family assignments for the Higgs fields given in Eq. (3),
the above terms for the Yukawa superpotential are uniquely obtained if
we extend the following $U(1) \times Z_2 \times Z_2$ assignments to the 
fermions:

$$\begin{array}{ll}
        {\bf 16}_3(-{1\over{2}}^{++}), \quad & {\bf 16}_2(\left[-{1\over{2}}
                + p\right]^{++}),\\[0.1in]
        {\bf 16}(-{1\over{2}}^{++}), \quad & {\bf \overline{16}}
                ({1\over{2}}^{++})\\[0.1in]
        {\bf 10}(-p^{-+}), \quad & {\bf 10'}(p^{++})\\
\end{array} \eqno(10)$$

The desired 22, 23, 32 and 33 entries in the Dirac matrices of Eq. (6)
are then obtained with the Yukawa interactions in Eq. (9) by integrating 
out the superheavy fermions introduced above.  The relevant diagrams are 
pictured in Fig. 1 where the asymmetrical nature of the contributions
is readily apparent.

In summary, we have shown that with the minimal set of $SO(10)$ Higgs fields 
introduced in Eq. (1) to solve the doublet-triplet splitting problem, 
fermion mass matrices can be constructed which explain well the known
quark mass and mixing data and lead to the suggestion of large $\nu_{\mu} -
\nu_{\tau}$ mixing responsible for the atmospheric neutrino anomaly.  Unlike
previous studies, this large neutrino mixing arises not from a large 
hierarchy in the right-handed Majorana matrix but rather as a result of 
the skewed spinor ${\bf 16'}$ Higgs and antisymmetrical $B - L$ adjoint 
${\bf 45}$ contributions to the Dirac matrices.\\

This work was supported in part by the Department of Energy Grant 
Nos. DE-FG02-91ER-40626, and DE-FG02-90ER-40542.  CHA thanks the Fermilab
Theoretical Physics Department for its kind hospitality.
%
%

%
%
%
\vspace*{0.5in}
Fig. 1.\ Diagrams that generate the 33, 23, 32 and 22 entries in the quark and 
lepton mass matrices of Eq. (6).  The second diagram of the 23 entry 
appears only for the down quark mass matrix.  A similar diagram in reverse
order would appear for the 32 entry of the charged lepton mass matrix.
\newpage
%
\vspace*{1in}
\begin{picture}(200,100)(35,0)
\vspace*{0.5in}
\thicklines
\put(120,144){\vector(1,0){45}}
\put(165,144){\line(1,0){45}}
\put(300,144){\vector(-1,0){45}}
\put(210,144){\line(1,0){45}}
\put(210,72){\vector(0,1){36}}
\put(210,108){\line(0,1){36}}
\put(35,162){${\bf 33:}$}
\put(157,162){${\bf 16_3}$}
\put(247,162){${\bf 16_3}$}
\put(220,100){${\bf T_1}$}
\end{picture}\\[0.8in]
%
%
\begin{picture}(200,100)(35,0)
\vspace*{0.5in}
\thicklines
\put(105,144){\vector(1,0){30}}
\put(135,144){\line(1,0){30}}
\put(165,144){\line(1,0){30}}
\put(225,144){\vector(-1,0){30}}
\put(225,144){\vector(1,0){30}}
\put(255,144){\line(1,0){30}}
\put(285,144){\line(1,0){30}}
\put(345,144){\vector(-1,0){30}}
\put(165,72){\vector(0,1){36}}
\put(165,108){\line(0,1){36}}
\put(225,72){\line(0,1){36}}
\put(225,144){\vector(0,-1){36}}
\put(285,72){\vector(0,1){36}}
\put(285,108){\line(0,1){36}}
\put(50,162){${\bf 23:}$}
\put(127,162){${\bf 16_2}$}
\put(187,162){${\bf 16}$}
\put(247,162){${\bf \overline{16}}$}
\put(307,162){${\bf 16_3}$}
\put(175,100){${\bf T_1}$}
\put(235,100){${\bf P}$}
\put(295,100){${\bf A}$}
\end{picture}\\[0.5in]
\begin{picture}(200,100)(-5,0)
\vspace*{0.5in}
\thicklines
\put(60,144){\vector(1,0){30}}
\put(90,144){\line(1,0){30}}
\put(120,144){\line(1,0){30}}
\put(180,144){\vector(-1,0){30}}
\put(180,144){\vector(1,0){30}}
\put(210,144){\line(1,0){30}}
\put(240,144){\line(1,0){30}}
\put(300,144){\vector(-1,0){30}}
\put(120,72){\vector(0,1){36}}
\put(120,108){\line(0,1){36}}
\put(180,72){\line(0,1){36}}
\put(180,144){\vector(0,-1){36}}
\put(240,72){\vector(0,1){36}}
\put(240,108){\line(0,1){36}}
\put(82,162){${\bf \overline{5}(16_2)}$}
\put(142,162){${\bf 5(10)}$}
\put(202,162){${\bf \overline{5}(10')}$}
\put(262,162){${\bf 10(16_3)}$}
\put(124,100){${\bf 1(C)}$}
\put(184,100){${\bf 1(\frac{\overline{C} C}{M_P})}$}
\put(244,100){${\bf \overline{5}(C')}$}
\end{picture}\\[0.5in]
\begin{picture}(200,100)(35,0)
\vspace*{0.5in}
\thicklines
\put(105,144){\vector(1,0){30}}
\put(135,144){\line(1,0){30}}
\put(165,144){\line(1,0){30}}
\put(225,144){\vector(-1,0){30}}
\put(225,144){\vector(1,0){30}}
\put(255,144){\line(1,0){30}}
\put(285,144){\line(1,0){30}}
\put(345,144){\vector(-1,0){30}}
\put(165,72){\vector(0,1){36}}
\put(165,108){\line(0,1){36}}
\put(225,72){\line(0,1){36}}
\put(225,144){\vector(0,-1){36}}
\put(285,72){\vector(0,1){36}}
\put(285,108){\line(0,1){36}}
\put(50,162){${\bf 32:}$}
\put(127,162){${\bf 16_3}$}
\put(187,162){${\bf \overline{16}}$}
\put(247,162){${\bf 16}$}
\put(307,162){${\bf 16_2}$}
\put(175,100){${\bf A}$}
\put(235,100){${\bf P}$}
\put(295,100){${\bf T_1}$}
\end{picture}\\[-0.5in]
\begin{picture}(200,1)(0,0)
%
\put(12,1){${\bf 22:}$ \qquad\ (None)}
\end{picture}\\

\newpage
\begin{references}
%
\bibitem{BR}   S.M. Barr and S. Raby, Phys. Rev. Lett. {\bf 79}, 4748 (1997).

\bibitem{ABB}  C.H. Albright, K.S. Babu and S.M. Barr, Report NO. 
	FERMILAB-Pub-98/052-T, IASSNS-HEP-98-14, BA-98-06.

\bibitem{Maj}   B. Brahmachari and R.N. Mohapatra,hep-ph/9710371;
        J. Sato and T. Yanagida, hep-ph/9710516;
        M. Drees, S. Pakvasa, X. Tata, and T. ter Veldhuis, hep-ph/9712392;
        M. Bando, T. Kugo, and K. Yoshioka hep-ph/9710417.

\bibitem{DW}   S. Dimopoulos and F. Wilczek, Report No. NSF-ITP-82-07 (1981),
        in {\it The unity of fundamental interactions}, Proceedings of
        the 19th Course of the International School of Subnuclear Physics,
        Erice, Italy, 1981, ed. A. Zichichi (Plenum Press, New York, 1983).

\bibitem{GJ} H. Georgi and C. Jarlskog, Phys. Lett. {\bf B86}, 297
        (1979).

\bibitem{atm}   K.S. Hirata {\it et al.}, Phys. Lett. B {\bf 205},
        416 (1988); K.S. Hirata {\it et al.}, Phys. Lett. B {\bf
        280}, 146 (1992); Y. Fukuda {\it et al.}, Phys. Lett. B {\bf
        335}, 237 (1994); D. Caspar {\it et al.}, Phys. Rev. Lett.
        {\bf 66}, 2561 (1991); R. Becker-Szendy {\it et al.},
        Phys. Rev. D {\bf 46}, 3720 (1992); Nucl. Phys. B
        (Proc. Suppl.) {\bf 38}, 331 (1995); T. Kafka, Nucl.
        Phys. B (Proc. Suppl.) {\bf 35}, 427 (1994); M. Goodman,
        {\it ibid.} {\bf 38}, 337 (1995); W.W.M. Allison {\it et al.},
        Phys. Lett. B {\bf 391}, 491 (1997).
\end{references}
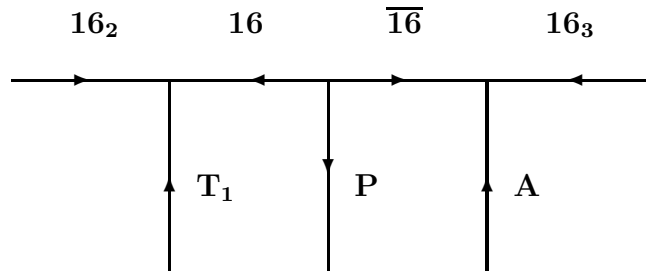
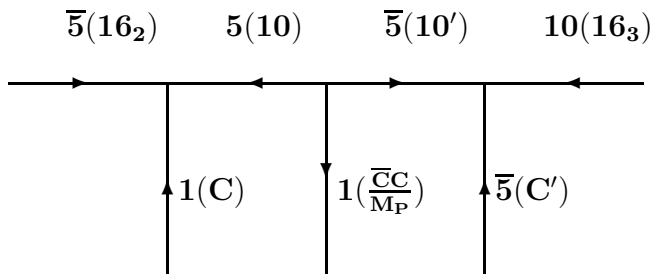
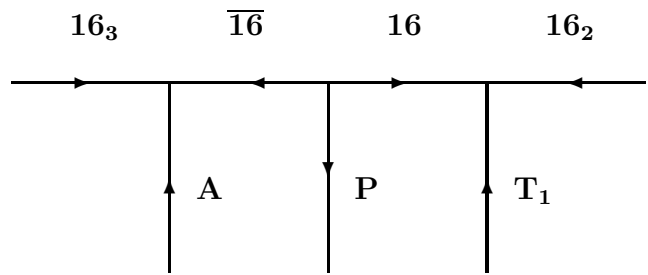
\end{document}